\documentclass[12pt]{article}
\usepackage{amsmath,amsfonts,amssymb}

\begin{document}
\thispagestyle{empty}

\def\theequation{\arabic{section}.\arabic{equation}}
\def\a{\alpha}
\def\b{\beta}
\def\g{\gamma}
\def\d{\delta}
\def\dd{\rm d}
\def\e{\epsilon}
\def\ve{\varepsilon}
\def\z{\zeta}
\def\B{\mbox{\bf B}}\def\cp{\mathbb {CP}^3}

\newcommand{\h}{\hspace{0.5cm}}

\begin{titlepage}
\vspace*{1.cm}
\renewcommand{\thefootnote}{\fnsymbol{footnote}}
\begin{center}
{\Large \bf Finite-size giant magnons on $\eta$-deformed $AdS_5\times S^5$}
\end{center}
\vskip 1.2cm \centerline{\bf Changrim  Ahn and Plamen Bozhilov
\footnote{On leave from Institute for Nuclear Research and Nuclear
Energy, Bulgarian Academy of Sciences, Bulgaria.}}

\vskip 10mm

\centerline{\sl Department of Physics} \centerline{\sl Ewha Womans
University} \centerline{\sl DaeHyun 11-1, Seoul 120-750, S. Korea}
\vspace*{0.6cm} \centerline{\tt ahn@ewha.ac.kr,
bozhilov@inrne.bas.bg}

\vskip 20mm

\baselineskip 18pt

\begin{center}
{\bf Abstract}
\end{center}
We consider strings moving in the $R_t\times S^3_\eta$ subspace of the
$\eta$-deformed $AdS_5\times S^5$ and obtain a class of solutions depending on several parameters.
They are characterized by the string energy and two angular momenta.
Finite-size dyonic giant magnon belongs to this class of solutions.
Further on, we restrict ourselves to the case of giant magnon with one nonzero angular momentum,
and obtain the leading finite-size correction to the dispersion relation.

\end{titlepage}
\newpage
\baselineskip 18pt

\def\nn{\nonumber}
\def\tr{{\rm tr}\,}
\def\p{\partial}
\newcommand{\non}{\nonumber}
\newcommand{\bea}{\begin{eqnarray}}
\newcommand{\eea}{\end{eqnarray}}
\newcommand{\bde}{{\bf e}}
\renewcommand{\thefootnote}{\fnsymbol{footnote}}
\newcommand{\be}{\begin{eqnarray}}
\newcommand{\ee}{\end{eqnarray}}

\vskip 0cm

\renewcommand{\thefootnote}{\arabic{footnote}}
\setcounter{footnote}{0}

\setcounter{equation}{0}
\section{Introduction}
In the recent years important progress has been made in the field of $AdS/CFT$ duality \cite{AdS/CFT}
(for overview see \cite{RO}).
The main achievements are due to the discovery of
integrable structures on both sides of the correspondence.

The most developed case is the correspondence
between strings moving in $AdS_5\times S^5$ and $\mathcal{N}=4$ SYM
in four dimensions. The so-called $\gamma$-deformation of $AdS_5\times S^5$
has been proposed in \cite{LM05}. It was shown in \cite{F05} that this deformation
is still integrable for real $\gamma$ (known as $\beta$- or $TsT$-deformation).

A new integrable deformation of the type IIB $AdS_5\times S^5$ superstring action,
depending on one real parameter $\eta$, has been found recently in \cite{DMV0913}.
The bosonic part of the superstring sigma model Lagrangian on this
$\eta$-deformed background was determined in \cite{ABF1312}.
Then the authors of \cite{ABF1312} used it to compute the perturbative
$S$-matrix of bosonic particles in the model.

Interesting new developments were made in \cite{ALT0314}.
There the spectrum of a string moving on $\eta$-deformed $AdS_5\times S^5$ is considered.
This is done by treating the corresponding worldsheet theory as integrable field theory.
In particular, it was found that the dispersion relation for the infinite-size
giant magnons \cite{HM06} on this background, in the large string tension limit $g\to\infty$ is given by
\bea\label{14} E=\frac{2g\sqrt{1+\tilde{\eta}^2}}{\tilde{\eta}} \mbox{arcsinh}\left(\tilde{\eta} \sin\frac{p}{2}\right),\eea
where $\tilde{\eta}$ is related to the deformation parameter $\eta$ according to
\bea\label{ek} \tilde{\eta}=\frac{2 \eta}{1-\eta^2}.\eea
Here, we are going to extend the result (\ref{14}) to the case of finite-size giant magnons.

The paper is organized as follows. In Sec.2 we give the bosonic part of the string Lagrangian on
$\eta$-deformed $AdS_5\times S^5$ found in \cite{ABF1312} and extract from it the background fields.
Then in Sec.3, we obtain the exact solutions
for the finite-size dyonic giant magnon coordinates, the corresponding conserved charges
and the angular difference along one of the isometric coordinates on the deformed sphere $S^3_\eta$
\footnote{This angular difference is identified with the momentum of the magnon excitations in the dual spin chain.}.
In Sec.4 we find the dispersion relation for the giant magnons with one nonzero angular momentum,
including the leading finite-size effect on it.
Sec.5 is devoted to our concluding remarks.

\setcounter{equation}{0}
\section{String Lagrangian and background fields}
The bosonic part of the string Lagrangian $\mathcal{L}$ on the
$\eta$-deformed $AdS_5\times S^5$ found in \cite{ABF1312} is given by a sum of the Lagrangians $\mathcal{L}_a$ and $\mathcal{L}_s$,
for the $AdS$ and sphere subspaces. Since there is nonzero $B$-field on both subspaces,
which leads to the appearance of Wess-Zumino terms,
these Lagrangians can be further decomposed as
\bea\label{Leta} \mathcal{L}_a=L_a^g+L_a^{WZ},\h \mathcal{L}_s=L_s^g+L_s^{WZ},\eea
where the superscript ``g'' is related to the dependence on the background metric.
The explicit expressions for the Lagrangians in (\ref{Leta}) are as follows \cite{ABF1312}

\bea\label{Lag} L_a^g &=&-\frac{T}{2}\gamma^{\alpha\beta}
\left[-\frac{(1+\rho^2)\p_\alpha t \p_\beta t} {1-\tilde{\eta}^2 \rho^2}+
\frac{\p_\alpha \rho \p_\beta \rho} {(1+\rho^2)(1-\tilde{\eta}^2 \rho^2)}+
\frac{\rho^2 \p_\alpha \zeta \p_\beta \zeta}{1+\tilde{\eta}^2 \rho^4 \sin^2\zeta} \right.
\\ \nn && \left. +\frac{\rho^2 \cos^2 \zeta \ \p_\alpha \psi_1 \p_\beta \psi_1}
{1+\tilde{\eta}^2 \rho^4 \sin^2\zeta}+\rho^2 \sin^2\zeta \ \p_\alpha \psi_2 \p_\beta \psi_2\right],\eea

\bea\label{LaWZ} L_a^{WZ}=\frac{T}{2}\tilde{\eta} \ \epsilon^{\alpha\beta}
\frac{\rho^4 \sin 2\zeta}{1+\tilde{\eta}^2 \rho^4 \sin^2\zeta}\ \p_\alpha\psi_1 \p_\beta\zeta,\eea

\bea\label{sag} L_s^g &=&-\frac{T}{2}\gamma^{\alpha\beta}
\left[\frac{(1-r^2)\p_\alpha \phi \p_\beta \phi} {1+\tilde{\eta}^2 r^2}+
\frac{\p_\alpha r \p_\beta r} {(1-r^2)(1+\tilde{\eta}^2 r^2)}+
\frac{r^2 \p_\alpha \xi \p_\beta \xi}{1+\tilde{\eta}^2 r^4 \sin^2\xi} \right.
\\ \nn && \left. +\frac{r^2 \cos^2 \xi \ \p_\alpha \phi_1 \p_\beta \phi_1}
{1+\tilde{\eta}^2 r^4 \sin^2\xi}+r^2 \sin^2\xi \ \p_\alpha \phi_2 \p_\beta \phi_2\right],\eea

\bea\label{LsWZ} L_s^{WZ}=-\frac{T}{2}\tilde{\eta} \ \epsilon^{\alpha\beta}
\frac{r^4 \sin 2\xi}{1+\tilde{\eta}^2 r^4 \sin^2\xi}\ \p_\alpha\phi_1 \p_\beta\xi,\eea
where we introduced the notation
\bea\label{T} T=g\sqrt{1+\tilde{\eta}^2}.\eea

Comparing (\ref{Lag})-(\ref{LsWZ}) with the Polyakov string Lagrangian,
one can extract the components of the background fields. They are given by
\bea\label{bfsa} && g_{tt}=-\frac{1+\rho^2}{1-\tilde{\eta}^2 \rho^2},\h
g_{\rho\rho}=\frac{1}{(1+\rho^2)(1-\tilde{\eta}^2 \rho^2)},\h g_{\zeta\zeta}= \frac{\rho^2}{1+\tilde{\eta}^2 \rho^4
\sin^2 \zeta}
\\ \nn && g_{\psi_1 \psi_1}= \frac{\rho^2\cos^2\zeta}{1+\tilde{\eta}^2 \rho^4\sin^2 \zeta},\h
g_{\psi_2 \psi_2}=\rho^2 \sin^2\zeta,\h b_{\psi_1\zeta}=\tilde{\eta} \frac{\rho^4 \sin 2\zeta}
{1+\tilde{\eta}^2 \rho^4\sin^2 \zeta}.\eea

\bea\label{bfss} && g_{\phi\phi}=\frac{1-r^2}{1+\tilde{\eta}^2 r^2},\h g_{rr}=\frac{1}{(1-r^2)(1+\tilde{\eta}^2 r^2)}
,\h g_{\xi\xi}=\frac{r^2}{1+\tilde{\eta}^2 r^4 \sin^2\xi}
\\ \nn && g_{\phi_1\phi_1}= \frac{r^2 \cos^2 \xi}{1+\tilde{\eta}^2 r^4 \sin^2\xi}.\h
g_{\phi_2\phi_2}= r^2 \sin^2\xi,\h b_{\phi_1\xi}=-\tilde{\eta} \frac{r^4 \sin 2\xi}{1+\tilde{\eta}^2 r^4 \sin^2\xi}.\eea

Since we are going to consider giant magnon solutions, we restrict ourselves to the $R_t\times S^3_\eta$ subspace,
which corresponds to the following choice in $AdS_\eta$
\bea\nn \rho=0,\h \zeta=0,\h \psi_1=\psi_2=0 \ \Rightarrow\ b_{\psi_1\zeta}=0.\eea
On $S^5_\eta$ we first introduce the angle $\tilde{\theta}$ in the following way
\bea\nn r=\sin\tilde{\theta},\eea
which leads to
\bea\nn ds^2_{S^5_\eta}&=& \frac{\cos^2 \tilde{\theta}}{1+\tilde{\eta}^2 \sin^2 \tilde{\theta}}\ d\phi^2
+\frac{d \tilde{\theta}^2}{1+\tilde{\eta}^2 \sin^2 \tilde{\theta}}
+\frac{\sin^2 \tilde{\theta}}{1+\tilde{\eta}^2 \sin^4 \tilde{\theta}\sin^2 \xi}\ d \xi^2
\\ \nn &&+\frac{\sin^2 \tilde{\theta} \cos^2 \xi}{1+\tilde{\eta}^2 \sin^4 \tilde{\theta}\sin^2 \xi}\ d \phi_1^2
+\sin^2 \tilde{\theta}\ \sin^2 \xi\ d \phi_2^2,\eea
\bea\nn b_{\phi_1\xi}=-\tilde{\eta} \frac{\sin^4\tilde{\theta}
\sin 2\xi}{1+\tilde{\eta}^2 \sin^4\tilde{\theta} \sin^2\xi}.\eea
Now, to go to $S^3_\eta$, we can safely set $\phi=0$, $\tilde{\theta}=\frac{\pi}{2}$
(we also exchange $\phi_1$ and $\phi_2$ and replace $\xi$ with $\theta$).
Thus, the background seen by the string moving in the $R_t\times S^3_\eta$ subspace
can be written as
\bea\nn &&g_{tt}=-1,\h g_{\phi_1\phi_1}=\sin^2\theta,\h g_{\phi_2\phi_2}=\frac{\cos^2\theta}{1+\tilde{\eta}^2 \sin^2\theta},
\\ \label{fb} &&g_{\theta\theta}=\frac{1}{1+\tilde{\eta}^2 \sin^2\theta},\h b_{\phi_2\theta}=
-\tilde{\eta} \frac{\sin 2\theta}{1+\tilde{\eta}^2 \sin^2\theta}.\eea

\setcounter{equation}{0}
\section{Exact results}
Here and further on we will work in conformal gauge when $\gamma^{\alpha\beta}=diag(-1,1)$ and
the string Lagrangian and Virasoro constraints have the form
\bea\label{Ls} &&L_s=\frac{T}{2}\left(G_{00}-G_{11}+2B_{01}\right),
\\ \label{Vir} &&G_{00}+G_{11}=0,\h G_{01}=0.\eea
Above
\bea\nn G_{\alpha\beta}=g_{MN}\p_\alpha X^M \p_\beta X^N,\h
B_{\alpha\beta}=b_{MN}\p_\alpha X^M \p_\beta X^N\eea
are the fields induced on the string worldsheet.
For the case under consideration
\bea\nn X^M=\left(t,\phi_1,\phi_2,\theta\right).\eea
The corresponding nonzero components of $g_{MN}$ and $b_{MN}$ are given
in (\ref{fb}).

Now, we impose the following ansatz for the string embedding
\bea\label{A} t(\tau,\sigma)=\kappa \tau,\h \phi_i(\tau,\sigma)=\omega_i \tau+F_i(\xi),
\h \theta(\tau,\sigma)=\theta(\xi),\h \xi=\alpha\sigma+\beta\tau ,\h i=1,2,\eea
where $\tau$ and $\sigma$ are the string world-sheet coordinates,
$F_i(\xi)$, $\theta(\xi)$ are arbitrary functions of $\xi$,
and $\kappa, \omega_i, \alpha, \beta$ are parameters.

Replacing (\ref{A}) into (\ref{Ls}) one finds the following solutions of the equations of motion
for $\phi_i(\tau,\sigma)$ (we introduced the notation $\chi\equiv \cos^2\theta$)
\bea\label{f1} &&\phi_1(\tau,\sigma)=\omega_1 \tau+\frac{1}{\alpha^2-\beta^2}\int d\xi\left(\frac{C_1}{1-\chi}
+\beta \omega_1\right),
\\\label{f2} &&\phi_2(\tau,\sigma)=\omega_2 \tau+\frac{1}{\alpha^2-\beta^2}\int d\xi \left[\frac{(1+\tilde{\eta}^2)C_2}{\chi}
+\beta \omega_2-\tilde{\eta}^2 C_2\right],\eea
where $C_1$, $C_2$ are integration constants.

By using (\ref{f1}), (\ref{f2}), one can show that the Virasoro constraints (\ref{Vir}) take the form
\bea\label{V1} \left(\frac{d \chi}{d \xi}\right)^2 &=&
\frac{4 \chi(1-\chi)\left[1+\tilde{\eta}^2(1-\chi)\right]}{(\alpha^2-\beta^2)^2}
\left[(\alpha^2+\beta^2)\kappa^2-\frac{C_1^2}{1-\chi}-C_2^2\frac{1+\tilde{\eta}^2(1-\chi)}
{\chi}\right.
\\ \nn &&-\left.\alpha^2 \omega_1^2(1-\chi)-
\alpha^2 \omega_2^2 \frac{\chi}{1+\tilde{\eta}^2(1-\chi)}\right],\eea
\bea\label{V2} \omega_1 C_1+\omega_2 C_2+\beta \kappa^2=0.\eea
Next, we solve (\ref{V2}) with respect to $C_1$ and replace the solution into (\ref{V1}).
The result is
\bea\label{cpf} \left(\frac{d \chi}{d \xi}\right)^2=
\frac{4}{(\alpha^2-\beta^2)^2}\alpha^2\tilde{\eta}^2\omega_1^2
(\chi_\eta-\chi)(\chi_p-\chi)(\chi-\chi_m)(\chi-\chi_n),\eea
where
\bea\label{roots1} \chi_\eta+\chi_p+\chi_m+\chi_n=
-\frac{\alpha^2\left[\omega_2^2-\omega_1^2+\tilde{\eta}^2(\kappa^2-3 \omega_1^2)
\right]+\tilde{\eta}^2\beta^2 \kappa^2+\tilde{\eta}^4 C_2^2}{\alpha^2\tilde{\eta}^2\omega_1^2},\eea
\bea\label{roots2} &&\chi_p \chi_\eta+(\chi_p + \chi_\eta)\chi_n+
\chi_m (\chi_p + \chi_\eta +\chi_n) =
\\ \nn && \frac{1}{\tilde{\eta}^2\alpha^2 \omega_1^4}
\left\{\beta^2\kappa^2\left[\tilde{\eta}^2(\kappa^2-2\omega_1^2)-\omega_1^2\right]
+2C_2\beta\tilde{\eta}^2\kappa^2\omega_2\right.
\\ \nn &&+\left. \alpha^2\omega_1^2\left[\left(2+3\tilde{\eta}^2\right)\omega_1^2
-\omega_2^2-\left(1+2\tilde{\eta}^2\right)\kappa^2\right]\right.
\\ \nn &&+\left. C_2^2\tilde{\eta}^2\left(\omega_2^2-\left(2+3\tilde{\eta}^2\right)\omega_1^2\right)\right\},\eea
\bea\label{roots3} &&\chi_m\chi_n\chi_p+\chi_m\chi_n\chi_\eta+\chi_m\chi_p\chi_\eta+\chi_n\chi_p\chi_\eta=
\\ \nn &&-\frac{1+\tilde{\eta}^2}{\tilde{\eta}^2\alpha^2 \omega_1^4}
\left[C_2^2(1+3\tilde{\eta}^2)\omega_1^2-2C_2\beta\kappa^2\omega_2-C_2^2\omega_2^2
-(\kappa^2-\omega_1^2)(\beta^2\kappa^2-\alpha^2\omega_1^2)\right],\eea
\bea\label{roots4} \chi_m\chi_n\chi_p\chi_\eta= -\frac{C_2^2(1+\tilde{\eta}^2)^2}{\tilde{\eta}^2\alpha^2 \omega_1^2}.\eea

The solution $\xi(\chi)$ of (\ref{cpf}) is
\bea\label{sxi} &&\xi(\chi) =
\frac{\alpha^2-\beta^2}{\tilde{\eta}\alpha\omega_1\sqrt{(\chi_\eta-\chi_m)(\chi_p-\chi_n)}}\times
\\ \nn
&&\mathbf{F}\left(\arcsin\sqrt{\frac{(\chi_\eta-\chi_m)(\chi_p-\chi)}{(\chi_p-\chi_m)(\chi_\eta-\chi)}},
\frac{(\chi_p-\chi_m)(\chi_\eta-\chi_n)}{(\chi_\eta-\chi_m)(\chi_p-\chi_n)}\right),\eea
where $\mathbf{F}$ is the incomplete elliptic integral of first kind and
\bea\nn \chi_\eta >\chi_p>\chi>\chi_m>\chi_n.\eea
Inverting $\xi(\chi)$ to $\chi(\xi)$, one finds
\bea\label{chis} \chi(\xi)= \frac{\chi_\eta(\chi_p-\chi_n)\ \mathbf{DN}^2(x,m)
+(\chi_\eta-\chi_p)\chi_n}{(\chi_p-\chi_n)\ \mathbf{DN}^2(x,m)+\chi_\eta-\chi_p},\eea
where $\mathbf{DN}(x,m)$ is one of the Jacobi elliptic functions and
\bea\nn &&x=\frac{\tilde{\eta} \alpha \omega_1\sqrt{(\chi_\eta-\chi_m)(\chi_p-\chi_n)}}
{\alpha^2-\beta^2}\ \xi ,
\\ \nn &&m= \frac{(\chi_p-\chi_m)(\chi_\eta-\chi_n)}{(\chi_\eta-\chi_m)(\chi_p-\chi_n)}.\eea

By using (\ref{cpf}) we can find the explicit solutions for the isometric angles $\phi_1$, $\phi_2$.
They are given by
\bea\label{f1s} &&\phi_1(\tau,\sigma)= \omega_1 \tau+\frac{1}{\tilde{\eta}\alpha\omega_1^2(\chi_\eta-1)
\sqrt{(\chi_\eta-\chi_m)(\chi_p-\chi_n)}} \times
\\ \nn &&\Bigg\{\Big[\beta\left(\kappa^2+\omega_1^2(\chi_\eta-1)+C_2 \omega_2\right)
\Big]\ \mathbf{F}\left(\arcsin\sqrt{\frac{(\chi_\eta-\chi_m)(\chi_p-\chi)}{(\chi_p-\chi_m)(\chi_\eta-\chi)}},
m\right)
\\ \nn &&-\frac{(\chi_\eta-\chi_p)(\beta\kappa^2+C_2 \omega_2)}{1-\chi_p}\
\mathbf{\Pi}\left(\arcsin\sqrt{\frac{(\chi_\eta-\chi_m)(\chi_p-\chi)}{(\chi_p-\chi_m)(\chi_\eta-\chi)}}
,-\frac{(\chi_\eta-1)(\chi_p-\chi_m)}{(1-\chi_p)(\chi_\eta-\chi_m)},m\right)\Bigg\},\eea
where $\mathbf{\Pi}$ is the incomplete elliptic integral of third kind.

\bea\label{f2s} &&\phi_2(\tau,\sigma)=\omega_2\tau +\frac{1}{\tilde{\eta}\alpha\omega_1\chi_\eta
\sqrt{(\chi_\eta-\chi_m)(\chi_p-\chi_n)}}
 \times
\\ \nn &&\Bigg\{\Big[C_2\left(1-\tilde{\eta}^2(\chi_\eta-1)\right)+\beta \omega_2 \chi_\eta
\Big]\ \mathbf{F}\left(\arcsin\sqrt{\frac{(\chi_\eta-\chi_m)(\chi_p-\chi)}{(\chi_p-\chi_m)(\chi_\eta-\chi)}},
m\right)
\\ \nn &&+\frac{C_2(1+\tilde{\eta}^2)(\chi_\eta-\chi_p)}{\chi_p}\times
\\ \nn
&&\mathbf{\Pi}\left(\arcsin\sqrt{\frac{(\chi_\eta-\chi_m)(\chi_p-\chi)}{(\chi_p-\chi_m)(\chi_\eta-\chi)}}
,\frac{\chi_\eta(\chi_p-\chi_m)}{(\chi_\eta-\chi_m)\chi_p},m\right)\Bigg\}.\eea

Now, let us go to the computations of the conserved charges $Q_\mu$, i.e. the string energy $E_s$
and the two angular momenta $J_1$, $J_2$.
Starting with
\bea\nn Q_\mu=\int d\sigma\frac{\p \mathcal{L}}{\p \left(\p_\tau X^{\mu}\right)},\h X^{\mu}=(t,\phi_1,\phi_2),\eea
and applying the ansatz (\ref{A}), one finds
\bea\label{Esi} E_s=\frac{T}{\tilde{\eta}}\left(1-\frac{\beta^2}{\alpha^2}\right)
\frac{\kappa}{\omega_1} \int_{\chi_m}^{\chi_p}
\frac{d\chi}{\sqrt{(\chi_\eta-\chi)(\chi_p-\chi)(\chi-\chi_m)(\chi-\chi_n)}},\eea
\bea\label{J1i} J_1&=&\frac{T}{\tilde{\eta}} \left[\left(1-\frac{\beta(\beta\kappa^2+C_2\omega_2)}{\alpha^2\omega_1^2}\right)
\int_{\chi_m}^{\chi_p}
\frac{d\chi}{\sqrt{(\chi_\eta-\chi)(\chi_p-\chi)(\chi-\chi_m)(\chi-\chi_n)}}\right.
\\ \nn &&-\left.\int_{\chi_m}^{\chi_p}
\frac{\chi d\chi}{\sqrt{(\chi_\eta-\chi)(\chi_p-\chi)(\chi-\chi_m)(\chi-\chi_n)}}\right],\eea
\bea\nn J_2&=&\frac{T}{\tilde{\eta}^3}
\Bigg[\left(1+\frac{1}{\tilde{\eta}^2}\right)\frac{\omega_2}{\omega_1}
\int_{\chi_m}^{\chi_p}
\frac{d\chi}{\left(1+\frac{1}{\tilde{\eta}^2}-\chi\right)\sqrt{(\chi_\eta-\chi)(\chi_p-\chi)(\chi-\chi_m)(\chi-\chi_n)}}
\\ \label{J2i} &&-\left(\frac{\omega_2}{\omega_1}-\tilde{\eta}^2 \frac{\beta C_2}{\alpha^2\omega_1}\right)
\int_{\chi_m}^{\chi_p}
\frac{d\chi}{\sqrt{(\chi_\eta-\chi)(\chi_p-\chi)(\chi-\chi_m)(\chi-\chi_n)}}\Bigg].\eea

We will need also the expression for the angular difference $\Delta\phi_1$.
The computations give the following result
\bea\label{adi} \Delta\phi_1 &=&\frac{1}{\tilde{\eta}}
\Bigg[\frac{\beta}{\alpha}\int_{\chi_m}^{\chi_p}
\frac{d\chi}{\sqrt{(\chi_\eta-\chi)(\chi_p-\chi)(\chi-\chi_m)(\chi-\chi_n)}}
\\ \nn &&-\left(\frac{\beta\kappa^2}{\alpha\omega_1^2}+
\frac{\omega_2C_2}{\alpha\omega_1^2}\right)\int_{\chi_m}^{\chi_p}
\frac{d\chi}{(1-\chi)\sqrt{(\chi_\eta-\chi)(\chi_p-\chi)(\chi-\chi_m)(\chi-\chi_n)}}\Bigg].\eea
Solving the integrals in (\ref{Esi})-(\ref{adi}) and introducing the notations
\bea\label{not} v=-\frac{\beta}{\alpha},\ u=\frac{\omega_2}{\omega_1},
\ W=\frac{\kappa^2}{\omega_1^2},\ K_2=\frac{C_2}{\alpha\omega_1},\
\epsilon=\frac{(\chi_\eta-\chi_p)(\chi_m-\chi_n)}{(\chi_\eta-\chi_m)(\chi_p-\chi_n)},\eea
we finally obtain
\bea\label{Esf} E_s=\frac{2T}{\tilde{\eta}} \frac{(1-v^2)\sqrt{W}}{\sqrt{(\chi_\eta-\chi_m)(\chi_p-\chi_n)}}
\ \mathbf{K}(1-\epsilon),\eea
\bea\label{J1f} J_1&=&\frac{2T}{\tilde{\eta}\sqrt{(\chi_\eta-\chi_m)(\chi_p-\chi_n)}}
\Bigg[\left(1-v^2W+K_2 u v-\chi_\eta\right)\ \mathbf{K}(1-\epsilon)
\\ \nn &&+(\chi_\eta-\chi_p)\ \mathbf{\Pi}\left(\frac{\chi_p-\chi_m}{\chi_\eta-\chi_m},
1-\epsilon\right)\Bigg],\eea
\bea\label{J2f} J_2&=& \frac{2T}{\tilde{\eta}^3\sqrt{(\chi_\eta-\chi_m)(\chi_p-\chi_n)}}
\Bigg\{\frac{\left(1+\frac{1}{\tilde{\eta}^2}\right)u}
{\left(1+\frac{1}{\tilde{\eta}^2}-\chi_\eta\right)}\times
\\ \nn &&\Bigg[\mathbf{K}(1-\epsilon)-\frac{\chi_\eta-\chi_p}{1+\frac{1}{\tilde{\eta}^2}-\chi_p}
\ \mathbf{\Pi}\left(\frac{(\chi_p-\chi_m)\left(1+\frac{1}{\tilde{\eta}^2}-\chi_\eta\right)}
{(\chi_\eta-\chi_m)\left(1+\frac{1}{\tilde{\eta}^2}-\chi_p\right)},1-\epsilon\right)\Bigg]
\\ \nn &&-\left(u+\tilde{\eta}^2 K_2 v\right)\ \mathbf{K}(1-\epsilon)\Bigg\},\eea
\bea\label{adf} \Delta\phi_1 &=& \frac{2}{\tilde{\eta}\sqrt{(\chi_\eta-\chi_m)(\chi_p-\chi_n)}}\times
\\ \nn &&\Bigg\{\frac{v W-K_2 u}{(\chi_\eta-1)(1-\chi_p)}\Bigg[(\chi_\eta-\chi_p)\
\mathbf{\Pi}\left(-\frac{(\chi_\eta-1)(\chi_p-\chi_m)}{(\chi_\eta-\chi_m)(1-\chi_p)},
1-\epsilon\right)
\\ \nn &&-(1-\chi_p)\ \mathbf{K}(1-\epsilon)\Bigg]-v \ \mathbf{K}(1-\epsilon)\Bigg\},\eea
where $\mathbf{K}$ and $\mathbf{\Pi}$ are the complete elliptic integrals of first and third kind.

\setcounter{equation}{0}
\section{Small $\epsilon$-expansions and dispersion relation}
In this section we restrict ourselves to the simpler case of giant magnons with one nonzero angular momentum.
To this end, we set the second isometric angle $\phi_2=0$.
From the solution (\ref{f2s}) it is clear that $\phi_2$ is zero when
\bea\nn \omega_2=C_2=0,\eea
or equivalently (see (\ref{not}))
\bea\nn u=K_2=0.\eea
Then it follows from (\ref{roots4}) that $\chi_n=0$ because $\chi_\eta>\chi_p>\chi_m>0$
for the finite-size case.
In addition, we express $\chi_m$ through the other parameters
in correspondence with (\ref{not})
\bea\nn \chi_m=\frac{\chi_\eta \chi_p}{\chi_\eta-(1-\epsilon)\chi_p}\ \epsilon.\eea
As a cosequence (\ref{roots1})-(\ref{roots3}) take the form
\bea\label{roots1r} \frac{(1-\epsilon)\chi_p^2-2\epsilon\chi_p\chi_\eta-\chi_\eta^2}{\chi_\eta-(1-\epsilon)\chi_p}
+3-(1+v^2)W+\frac{1}{\tilde{\eta}^2}=0,\eea
\bea\label{roots2r} \chi_p\chi_\eta+\frac{\epsilon\chi_p\chi_\eta(\chi_p+\chi_\eta)}{\chi_\eta-(1-\epsilon)\chi_p}
-\frac{2-(1+v^2)W+\left(3-\left(2+v^2(2-W)\right)W\right)\tilde{\eta}^2}{\tilde{\eta}^2}=0,\eea
\bea\label{roots3r} \frac{\epsilon\chi_p^2\chi_\eta^2}{\chi_\eta-(1-\epsilon)\chi_p}
-\frac{(1+\tilde{\eta}^2)(1-W)(1-v^2 W)}{\tilde{\eta}^2}=0.\eea

In order to obtain the leading finite-size effect on the dispersion relation,
we consider the limit $\epsilon\to 0$ in (\ref{roots1r})-(\ref{roots3r}) first.
We will use the following small $\epsilon$-expansions for the remaining parameters
\bea\label{chie} &&\chi_{\eta}= \chi_{\eta 0}+(\chi_{\eta 1}+\chi_{\eta 2}\log\epsilon)\epsilon
\\ \nn &&\chi_{p}= \chi_{p 0}+(\chi_{p 1}+\chi_{p 2}\log\epsilon)\epsilon,
\\ \nn &&v= v_0+(v_1+v_2\log\epsilon)\epsilon,
\\ \nn &&W= 1+W_1 \epsilon .\eea
Replacing (\ref{chie}) into (\ref{roots1r})-(\ref{roots3r}) and expanding in $\epsilon$
one finds the following solution of the resulting equations
\bea\label{chisol} &&\chi_{p 0}=1-v_0^2,\h \chi_{p 1}=1-v_0^2-2v_0 v_1-
\frac{(1-v_0^2)^2}{1+\tilde{\eta}^2 v_0^2},\h \chi_{p 2}=-2v_0 v_2,
\\ \nn &&\chi_{\eta 0}=1+\frac{1}{\tilde{\eta}^2},\h \chi_{\eta 1}=\chi_{\eta 2}=0,
\\ \nn &&W_1=-\frac{(1+\tilde{\eta}^2)(1-v_0^2)}{1+\tilde{\eta}^2 v_0^2}.\eea

Next, we expand $\Delta\phi_1$ in $\epsilon$ and impose the condition that
the resulting expression does not depend on $\epsilon$. After using (\ref{chisol}) this gives
\bea\label{df1} \Delta\phi_1=2\ \mbox{arccot}\left(v_0\sqrt{\frac{1+\tilde{\eta}^2}{1-v_0^2}}\right)
\eea
and two equations with solution
\bea\label{v1v2} v_1=\frac{v_0(1-v_0^2)\left[1-\log 16 +\tilde{\eta}^2
\left(2-v_0^2(1+\log 16 )\right)\right]}{4(1+\tilde{\eta}^2 v_0^2)},
\h v_2=\frac{1}{4}v_0(1-v_0^2).\eea
Solving (\ref{df1}) with respect to $v_0$ one finds
\bea\label{v0sol} v_0=\frac{\cot\frac{\Delta\phi_1}{2}}{\sqrt{\tilde{\eta}^2+\csc^2\frac{\Delta\phi_1}{2}}}.\eea

Now let us go to the $\epsilon$-expansion of the difference $E_s-J_1$.
Taking into account the solutions for the parameters, it can be written as
\bea\label{fr} E_s-J_1= 2 g \sqrt{1+\tilde{\eta}^2} \left[\frac{1}{\tilde{\eta}}
\mbox{arcsinh}\left(\tilde{\eta} \sin\frac{p}{2}\right)-\frac{(1+\tilde{\eta}^2)
\sin^3\frac{p}{2}}{4 \sqrt{1+\tilde{\eta}^2 \sin^2\frac{p}{2}}}\ \epsilon\right].\eea
where the expression for $\epsilon$ can be found from the expansion of $J_1$.
To the leading order, the result is
\bea\label{eps} \epsilon =16\ \exp\left[-\left(\frac{J_1}{g}
+\frac{2\sqrt{1+\tilde{\eta}^2}}{\tilde{\eta}}\mbox{arcsinh}\left(\tilde{\eta} \sin\frac{p}{2}\right)
\right)\sqrt{\frac{1+\tilde{\eta}^2\sin^2\frac{p}{2}}{\left(1+\tilde{\eta}^2\right)\sin^2\frac{p}{2}}}
\right].\eea
In writing (\ref{fr}), (\ref{eps}), we used (\ref{T}) and identified the angular difference
$\Delta\phi_1$ with the magnon momentum $p$ in the dual spin chain.

For $\epsilon=0$, (\ref{fr}) reduces to the dispersion relation for the infinite-size giant magnon
obtained in \cite{ALT0314} for the large $g$ case.
In the limit $\tilde{\eta}\to 0$, (\ref{fr}) gives the correct result
for the undeformed case found in \cite{AFZ06}.

\setcounter{equation}{0}
\section{Concluding Remarks}
Here we dealt with strings moving in the $R_t\times S^3_\eta$ subspace of the
$\eta$-deformed $AdS_5\times S^5$. The finite-size dyonic giant magnon solution is contained in
the string configurations we considered.

We derived the explicit exact solutions for the string coordinates and the
corresponding conserved charges. Then we restricted ourselves to the case of giant magnons
with one nonzero angular momentum and obtained the dispersion relation for them
including the leading finite-size effect on it.

It will be interesting to extend the result (\ref{fr}), (\ref{eps})
to the case of {\it dyonic} giant magnons. We will report on this soon.

Another possible direction of further investigation is to show that
(\ref{fr}), (\ref{eps}) can be reproduced by using the L\"uscher formula for the finite-size
effects on the dispersion relation and we are going to do that in the near future.

\section*{Acknowledgements}
This work was supported in part by the 
Research fund no. 1-2010-2469-001 by Ewha Womans University (CA), 
and the Brain Pool program (131S-1-3-0534) from the Ministry of Science, Ict and future
Planning.

\end{document}